\documentstyle[12pt]{article}
\begin{document}
\[ \]
\begin{center}
{\large \bf Dynamics of a Spherical Null Shell within\\[0.1cm] the
Distributional Formalism}
\[ \]
\[\]{\bf Samad Khakshournia\footnote {Email address:
skhakshour@aeoi.org.ir}}

Nuclear Research Center, Atomic Energy Organization of Iran,
Tehran, Iran
\[\]

{\bf Reza Mansouri\footnote {Email address: Mansouri@sharif.edu}}

Department of Physics, Sharif University of Technology, Tehran,
Iran.

and

Institute for studies in Physics and Mathematics. P.O.Box 5531,
Tehran, Iran.
\[ \]
\end{center}
\[ \]
\begin{abstract}

Dynamics of a null thin shell immersed in a generic spherically
symmetric spacetime is obtained within the distributional
formalism. It has been shown that the distributional formalism
leads to the same result as in the conventional formalism.
\end{abstract}
\newpage
\vspace{1.5cm}

\section{Introduction}
\hspace*{0.5cm}Many practical problems of general relativity and
cosmology involve idealized models constructed by gluing two
regions with different metrics across a hypersurface or thin shell
having a $\delta$-function singularity in its Riemann tensor due
to the discontinuity in the metric's transverse derivative across
the shell. The description of timelike(or spacelike) thin shells
is well known within general relativity since the outstanding work
of Israel[1]. Later, an extension of the Israel formalism to the
null or lightlike case was presented by Barrab$\grave{e}$s and
Israel[2]. Very recently, Poisson has introduced an user-friendly
reformulation of the Barrab$\grave{e}$s-Israel original work
together with an illustration of the formalism[3].\\
In general, one can distinguish two equivalent approaches to
describe thin shells or singular hypersurfaces as the boundary of
two glued manifolds. The well-known Israel formalism relates the
jump in the transverse curvature to properties of the singular
hypersurface. This formalism, being purely intrinsic, allows an
independent and arbitrary choice of coordinates at both sides of
the shell.\\
There is another formalism based on the distributional theory
requiring a preconstruction of a common set of coordinates
covering both sides of the shell making the four-metric continuous
across the shell[4]. In this approach, for the non-light-like
case, it has been shown that the singularity in the Ricci part of
the Riemann tensor is directly associated with the stress energy
tensor supported by the shell. The light-like case has recently
been developed by Nozari and Mansouri[5], and applied to the
spherically symmetric shell. They, however, have missed the
construction of a suitable set of coordinates to make the
four-metric continuous across the spherically symmetric layer, as
it is required by the distributional formalism. Our task is to
remedy this deficiency and incorporate an admissible continuous
coordinate system across the null layer leading to the correct
junction equations obtained in Ref. [2].\\
The plan of the paper is as follows. In section 2 we review
shortly the distributional formalism for null shells and give the
necessary formulae as a ready recipe to use. In section 3 we
consider spherical null-like shells as a simple example to examine
the efficiency of the distributional method.\\
 Conventions: we use the metric signature $(- + + +)$, and define
the Ricci tensor as $R_{\mu \nu} =\Gamma_{\mu \rho ,\nu}^{\rho}$
-...,. The Greek indices run from 0 to 3 and Latin indices $a,b$
from 1 to 3 but $A$ and $B$ takes only values 2 and 3. The square
brackets, $[F]$, are used to indicate the jump of any quantity $F$
across the layer. Terms proportional to $\delta$-function are
denoted by $\breve{F}$.
\section{Null-Shell Distributional Formalism}
\hspace*{0.5cm}Consider a space-time manifold $\cal M$ consisting
of overlapping domains $\cal M_{+}$ and $\cal M_{-}$ with metrics
$g_{\alpha \beta}^{+} (x_{+}^{\mu})$ and $g_{\alpha
\beta}^{-}(x_{-}^{\mu})$ in terms of independent disconnected
charts $x_{+}^{\mu}$ and $x_{-}^{\mu}$, respectively. The common
boundary of the domains is denoted by $\Sigma$ and taken to be
light-like. In other words, the Manifolds $\cal M_{+}$ and $\cal
M_{-}$ are glued together along the null hypersurface $\Sigma$.
Introducing a single chart $x^{\mu}$ called admissible coordinate
system that covers the overlap and reaches into both domains, we
write down the parametric equation of $\Sigma$ as $\Phi
(x^{\mu})=0$, where $\Phi$ is a smooth function[2]. The domains of
$\cal M$ in which $\Phi$ is positive or negative are contained in
$\cal M_{+}$ or $\cal M_{-}$, respectively. By applying the
coordinate transformations $x_{\pm}^{\mu}=x_{\pm}^{\mu} (x^{\nu})$
on the corresponding domains, a pair of metrics $g_{\alpha
\beta}^{+}(x^{\mu})$  and $g_{\alpha \beta}^{-}(x^{\mu})$ is
formed over $\cal M_{+}$ and $\cal M_{-}$ respectively, each
suitably smooth(say $C^{3}$). \\
The main step in the distributional approach is the definition of
a hybrid metric $g_{\alpha \beta}(x^{\mu})$ over $\cal M$ which
glues the metrics $g_{\alpha \beta}^{+}(x^{\mu})$ and $g_{\alpha
\beta}^{-}(x^{\mu})$ together continuously on $\Sigma$:
\begin{eqnarray}
g_{\alpha \beta}=g_{\alpha \beta}^{+}\theta (\Phi )+g_{\alpha
\beta}^ {-}\theta (-\Phi ),
\end{eqnarray}
where $\theta$ is the Heaviside step function and
\begin{eqnarray}
\left[ g_{\alpha \beta} (x^{\mu}) \right] =0.
\end{eqnarray}
We expect on $\Sigma$ the curvature and Ricci tensor to be
proportional to $\delta$ function. It follows from (1) and (2)
that the first derivative of $g_{\alpha \beta}$ is proportional
to the step function. The $\delta$ distribution can only occur in
the second derivative of the metric which enters linearly in the
expressions for curvature and Ricci tensor. So the only relevant
terms in the Ricci tensor are
\begin{eqnarray}
\check{R}_{\mu \nu} =\check{\Gamma}_{\mu \rho ,\nu}^{\rho} -
\check{\Gamma}_{\mu \nu , \rho}^{\rho}.
\end{eqnarray}
Using the metric in the form (1), we finally arrive at the
following expression for the components of the Ricci tensor
proportional to $\delta$ distribution [4]
\begin{eqnarray}
\check{R}_{\mu \nu}= \left( \frac{1}{2g} [g_{,\mu}]
\partial_{\nu} \phi - \left[ \Gamma_{\mu \nu}^{\rho} \right]
\partial_{\rho} \phi \right) \delta (\Phi ),
\end{eqnarray}
where $g$ is the determinant of the metric and the partial
derivatives are done with respect to the admissible coordinates
$x^{\mu}$.\\
The intrinsic coordinates of $\Sigma$ adapted to its null
generators are taken to be $\xi^{a}=(\eta,\theta^{A})$, with
$\eta$ being an arbitrary parameter( not necessarily affine on
either side of $\Sigma$) on the null generators of the
hypersurface and $\theta^{A}$ as labels of the generators. Now we
introduce tangent vectors $e_{a}^{\mu} = \frac{\partial
x^{\mu}}{\partial \xi^{a}}$, naturally segregated into a null
normal vector $n_{\mu}=\alpha^{-1}\partial_{\mu}\Phi$ that is also
tangent to the generators, and two space-like vectors
$e_{A}^{\mu}$ pointing in directions transverse to the
generators[3]
\begin{eqnarray}
n^{\mu}=(\frac{\partial x^{\mu}}{\partial
\eta})_{\theta^{A}}\equiv e_{\eta}^{\mu},\ \ \ \ \ e_{A}^{\mu} =
(\frac{\partial x^{\mu}}{\partial \xi^{A}})_{\eta}.
\end{eqnarray}
By construction, these vectors satisfy
$n^{\mu}n_{\mu}=0=n_{\mu}e_{A}^{\mu}$. We now complete the partial
basis $e_{a}^{\mu}$ by adding a transverse null vector $N^{\mu}$
with the following properties[3]
\begin{eqnarray}
N^{\mu}N_{\mu}=0,\ \ \ \   N_{\mu}n^{\mu}=-1,\ \ \ \
N_{\mu}e_{A}^{\mu}=0.
\end{eqnarray}
Therefore
\begin{eqnarray}
\alpha=-N^{\mu}\partial_{\mu} \Phi.
\end{eqnarray}
The intrinsic metric on $\Sigma$ may then be written as
\begin{eqnarray}
\gamma_{AB}=g_{\mu\nu}e_{A}^{\mu}e_{B}^{\nu},
\end{eqnarray}
and must be the same on the both sides of $\Sigma$. Following
jumps on $\Sigma$ turn out to be vanishing:
$[\gamma_{AB}]=[n^{\mu}]=[e_{A}^{\mu}]=[N^{\mu}]=[\alpha]=0$
expressed in the admissible coordinates $x^{\mu}$.\\
The energy-momentum tensor of the shell $\check{T}_{\mu \nu}$,
considered as a distribution, is given by [2-4]
\begin{eqnarray}
\check{T}_{\mu \nu}=|\alpha |S_{\mu \nu} \delta (\Phi ),
\end{eqnarray}
where $S_{\mu \nu}$ is the surface tensor of energy-momentum of
the shell expressed in the admissible coordinates $x^{\mu}$[3]:
\begin{eqnarray}
-\epsilon S^{\mu\nu}=\sigma
n^{\mu}n^{\nu}+j^{A}(n^{\mu}e_{A}^{\nu}+e_{A}^{\mu}n^{\nu})+p\gamma^{AB}e_{A}^{\mu}e_{B}^{\nu},
\end{eqnarray}
with $\epsilon=\frac{|\alpha|}{\alpha}$. The first term represents
a flow of matter along the null generators of the hypersurface,
and hence $\sigma$ represents a mass density. The second term
represents a flow of matter in the direction transverse to the
generators. Therefore, $j$ represents a current density.
Finally, the surface quantity $p$ represents an isotropic pressure.\\
Now we may write Einstein's field equation for the lightlike
hypersurface $\Sigma$ as[5]
\begin{equation}
\label{math:3.21} \breve{G}_{\mu\nu} = -\kappa\breve{T}_{\mu\nu}.
\end{equation}
Now taking into account Eq. (4) we define
\begin{equation}
\label{math:3.22} Q_{\mu\nu} = \left(\frac{1}{2g} [g_{,\mu}] \
\delta_{\nu}^{\rho}  - [\Gamma_{\mu\nu}^\rho] \right)n_{\rho}.
\end{equation}
Using Eq. (8) for the energy-momentum tensor we may write down
Eq.(11) as[5]
\begin{equation}
\label{math:3.23} Q_{\mu\nu} - \frac{1}{2}g_{\mu\nu}Q
=-\epsilon\kappa S_{\mu\nu},
\end{equation}
where $Q = Q_{\mu\nu}g^{\mu\nu}$, and $Q_{\mu\nu}$ is a tensor
with support on $\Sigma$. This so-called Sen equation, obtained in
the admissible coordinate system, describes the dynamics of null
surface layer $\Sigma$ within the distributional approach.

\section{Spherical Null shells}
\hspace*{0.5cm}To see the efficacy of the distributional method we
consider the situation in which the null shell is immersed in a
general spherical symmetric spacetime expressed in terms of the
Eddington retarded or advanced time $u$[2]:
\begin{equation}
\label{math:4.1} ds^{2}= -e^{\psi}du(f e^ {\psi}du + 2\zeta dr) +
r^{2}d\Omega^{2},
\end{equation}
where $\psi_{\pm}$ and $f_{\pm}$ are two arbitrary functions of
the coordinates $u_{\pm}$ and $r_{\pm}$. The sign factor $\zeta$
is $+1$ ($-1$) if $r$ increases (decreases) along a ray $u =
constant$, i.e., if the light cone $u = constant$ is expanding
(contracting). It is convenient to introduce the mass function
$m_{\pm}(u_{\pm},r_{\pm})$ defined as $f = 1-\frac{2m}{r}$.
Consider now a thin spherical shell whose history $\Sigma$ being a
light cone $u = constant$ splits the spacetime into the past- and
future-domains $\cal M_{-}$ and $\cal M_{+}$. Our aim is to glue
two spacetimes $\cal M_{-}$ and $\cal M_{+}$ along the
hypersurface $\Sigma$ using our distributional approach.\\
First we look for the admissible coordinate system
$x^{\mu}=(u,r,\theta,\phi)$ in which the parametric equation
describing the null shell is written as
\begin{equation}
\Phi(x^{\mu}) = u-R(r)=0.
\end{equation}
Now, we apply the following transformations to make the
four-metric continuous on the shell
\begin{eqnarray}
\begin{array}{cc} u_{-}=u,& \ \ \ \ u_{+}=A(u,r),\\ r_{-}=r,&
\ \ \ \ r_{+}=B(u,r). \end{array}
\end{eqnarray}
Carrying out the transformations and requiring the continuity of
the metric on $\Sigma$ according to Eq. (2), we obtain
\begin{eqnarray}
\left\{ \begin{array}{lr} U\equiv
f_{+}e^{2\psi_{+}}A_{,u}^{2}+2\zeta e^{\psi_{+}}A_{,u}B_{,u}
\stackrel{\Sigma}{=}f_{-}e^{2\psi_{-}},\\
 X\equiv f_{+}e^{2\psi_{+}}A_{,r}^{2}+2\zeta
 e^{\psi_{+}}A_{,r}B_{,r}\stackrel{\Sigma}{=}0,\\
 W\equiv f_{+}e^{2\psi_{+}}A_{,u}A_{,r}+\zeta e^{\psi_{+}}(A_{,u}B_{,r}+A_{,r}B_{,u})
 \stackrel{\Sigma}{=}\zeta e^{\psi_{-}},\\
 B(u,r)\stackrel{\Sigma}{=}r, \end{array} \right.
\end{eqnarray}
where $\stackrel{\Sigma}{=}$ means that both sides of the
equality are evaluated on $\Sigma$. Taking $\eta=\zeta r$ to be
the parameter on the null generators, from (5) the tangent
vectors are given:
\begin{equation}
n^{\mu}\partial_{\mu}=\zeta R_{,r}\partial_{u}+\zeta
\partial_{r},\ \ \
e_{\theta}^{\mu}\partial_{\mu}=\partial_{\theta},\ \
e_{\phi}^{\mu}\partial_{\mu}=\partial_{\phi}.
\end{equation}
According to Eq. (8) the shell's intrinsic two-metric is given by
\begin{equation}
ds^{2}_{\Sigma}=r^{2}(d\theta^{2}+\sin^{2}\theta d\phi^{2}).
\end{equation}
The continuity of  the induced metric on $\Sigma$ dictates the
following condition on $R(r)$
\begin{equation}
R_{,r}=-2\zeta \frac{e^{\psi_{-}}}{f_{-}}\Bigl|_{\Sigma}.
\end{equation}
Note that in the admissible coordinate $x^{\mu}$ constructed by the
transformations (16) the null hypersurface $\Sigma$ is given by
$fe^{\psi}du + 2\zeta dr=0$. The components of the transverse null
vector computed from (6) are
\begin{equation}
N_{\mu}dx^{\mu}=\frac{1}{2}f_{-}e^{\psi_{-}}du\Bigl|_{\Sigma}.
\end{equation}
Hence, using (7) we get $\alpha= e^{-\psi_{-}}$. From the last
equation of (17) we obtain
\begin{equation}
f_{-}B_{,r}-2\zeta e^{-\psi_{-}}B_{,u}\stackrel{\Sigma}{=}f_{-}.
\end{equation}
Now, using (22) the set of equations (17) can be solved for the
unknown functions $A_{u}$, $A_{r}$, $B_{u}$, and $B_{r}$:\\
\begin{eqnarray}
\left\{ \begin{array}{lr}A_{,u}\stackrel{\Sigma}{=}-e^{\psi_{-}-\psi_{+}},\\

A_{,r}\stackrel{\Sigma}{=}\frac{-2\zeta}{ f_{-}}e^{-\psi_{+}},\\

B_{,u}\stackrel{\Sigma}{=}\frac{1}{2}\zeta (f_{+}-f_{-})e^{\psi_{-}},\\

B_{,r}\stackrel{\Sigma}{=}\frac{f_{+}}{f_{-}}.
\end{array} \right.
\end{eqnarray}
The nonvanishing components of $Q_{\mu\nu}$ computed from Eq. (12)
are
\begin{eqnarray}
Q_{uu}=\frac{\zeta[f]}{r}e^{2\psi_{-}}-\zeta[\partial_{r}\psi]f_{-}e^{2\psi_{-}}\Bigl|_{\Sigma},
\end{eqnarray}
\begin{eqnarray}
Q_{rr} =4\zeta \frac{[f]}{rf_{-}^{2}}\Bigl|_{\Sigma},
\end{eqnarray}
\begin{eqnarray}
Q_{ur}=Q_{ru}=\frac{2[f]}{rf_{-}}e^{\psi_{-}}-[\partial_{r}\psi]e^{\psi_{-}}\Bigl|_{\Sigma},
\end{eqnarray}
where the derivatives in $[\partial_{r}\psi]$ are taken with
respect to the relevant radial coordinates $r_{\pm}$. Now, we can
immediately calculate $Q$ as
\begin{equation}
Q=2\zeta[\partial_{r}\psi]\Bigl|_{\Sigma}.
\end{equation}
The nonzero components of the surface energy tensor $S^{\mu\nu}$
are then calculated to be
\begin{equation}
S_{uu}= -\sigma e^{2\psi_{-}}\Bigl|_{\Sigma},
\end{equation}
\begin{equation}
S_{rr}= -4\sigma f_{-}^{-2}\Bigl|_{\Sigma},
\end{equation}
\begin{equation}
S_{ur}= -2\zeta\frac{\sigma}{f_{-}}e^{\psi_{-}}\Bigl|_{\Sigma},
\end{equation}
\begin{equation}
S_{\theta\theta}= -pr^{2}\Bigl|_{\Sigma},\ \ \ \
S_{\phi\phi}=\sin^{2}\theta S_{\theta\theta}.
\end{equation}
Finally, we obtain the following junction equation from the
non-angular components of Eq. (13)
\begin{equation}
\sigma=(-\zeta)\frac{[m]}{4\pi r^{2}}\Bigl|_{\Sigma}.
\end{equation}
Whereas the angular components of Eq. (13) yield
\begin{equation}
p=(-\zeta)\frac{1}{8\pi}[\partial_{r}\psi]\Bigl|_{\Sigma}.
\end{equation}
These are the same results as those obtained by C. Barrab\`{e}s
and W. Israel(see Eqs. (51) in Ref. [2]).
\section{Conclusion}
\hspace*{0.5cm}Direct application of the distributional formalism
to describe thin layers in general relativity requires a
preconstruction of spacetime coordinates that match continuously
on the shell to make the four-metric continuous. In consequence of
missing this point in Ref. [5], an incorrect junction equation
(Eq. (47) in Ref. [5]) was obtained. We have explicitly shown that
in the case of a null shell embedded in a generic spherically
symmetric geometry, it is possible to construct such an admissible
coordinate system covering both sides of shell, leading to the
same result as the common Israel formalism. This construction
allows us to apply the distributional formulation to treat null
surface layers in general relativity.

\end{document}